# The effects of spatiotemporal coherence on interferometric imaging

Seungwoo Shin[1,3], Kyoohyun Kim[1,4], KyeoReh Lee[1,5], SangYun Lee[1,6],
and YongKeun Park[1,2*]

[1]Department of Physics, Korea Advanced Institute of Science and Technology, Daejeon 34141, Republic of Korea
[2]Tomocube Inc., Daejeon 34051, Republic of Korea
[3]fire6516@kaist.ac.kr
[4]kkh2580@kaist.ac.kr
[5]kyeo@kaist.ac.kr
[6]prism@kaist.ac.kr
*Corresponding author, E-mail: yk.park@kaist.ac.kr, Tel: +82-42-350-2514, Fax: +82-42-350-2510

## Abstract

Illumination coherence plays a major role in various imaging systems, from microscopy, metrology, digital holography, optical coherence tomography, to ultrasound imaging. Here, we present a systematic study on the effects of degrees of spatiotemporal coherence of an illumination (DSTCI) on imaging quality. An optical field with arbitrary DSTCI was decomposed into wavelets with constituent spatiotemporal frequencies, and the effects on image quality were quantitatively investigated. The results show the synergistic effects on reduction of speckle noise when DSTCI is decreased. This study presents a method to systematically control DSTCI, and the result provides an essential reference on the effects of DSTCI on imaging quality. We believe that the presented methods and results can be implemented in various imaging systems for characterising and improving imaging quality.

## INTRODUCTION

The degree of coherence of a wave divulges a range of a predictable relative phase of the wave at a different time or position[1]. Degree of coherence is a fundamental concept in wave physics, and it has particularly important implications in imaging systems in such areas as ultrasound imaging[2], telescopy[3], interferometry[4], and biological and industrial applications[5]. Particularly in optics, numerous applications have been related to and affected by the degree of coherence of a light source, ranging from optical microscopy[6], speckle metrology[7], interferometric imaging[8], holography[9], optical coherence tomography[10], quantitative phase imaging[11, 12], to optical diffraction tomography[13, 14]. Although the use of a coherent illumination source such as in a laser ensures high radiance, well-defined wavelengths, and phase retrieval via interference, it inevitably suffers from unwanted diffraction effects called speckle noise or parasitic fringes[15]. Speckle noise originates from the superposition of a randomly varying optical field of waves caused by factors such as unwanted diffraction from dust particles, imperfect optical alignments, multiple reflections, and even laser oscillator cavities[15].

For decades, several sporadic studies have reported that the use of low coherent sources in either spatial[16-22] or temporal[23-30] domains can provide better interferometric imaging quality, mainly by reducing speckle noise or parasitic fringe patterns. Previous studies are summarised in Figure 1, with information about spatial and temporal coherence lengths of varying illuminations; the coherence length is a quantified value of the degree of coherence in either a spatial or temporal domain. For example, it is well known that the use of rotating diffusers can decrease spatial coherence length[16-21], and temporally low coherent sources, such as LEDs, can decrease temporal coherence lengths[23-30]. Recently, it was shown that the use of a random laser could significantly reduce the speckle noise in interferometric imaging while maintaining spectral radiance[21], which results in the decrease of speckle noise. Due to the importance of the degree of illumination coherence, the effect of suppressing speckle noise with spatially low coherent illumination has been investigated[31, 32]. Until now, however, due to the difficulty in systematic control of the degrees of spatiotemporal coherence of an illumination (DSTCI) in an interferometric imaging system, a quantitative investigation of the relationship between DSTCI and imaging quality has remained unexplored.

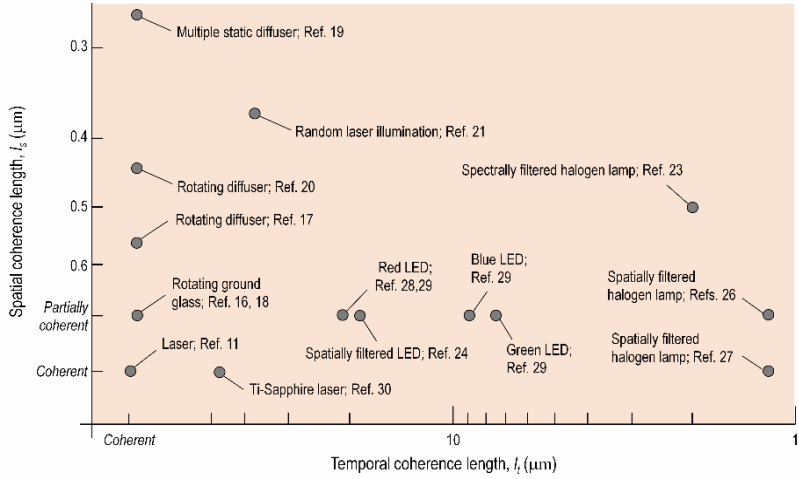

Figure 1. Preceding studies classified by degree of spatiotemporal coherence of the illumination. Reports in which the degree of coherence was not specified are classified as "partially coherent."

We experimentally investigated the relationship between DSTCI and interferometric imaging quality. To this purpose, we employed a quantitative phase imaging technique[11, 12] and have proposed an approach to control and analyse DSTCI. We measured optical field images with varying spatiotemporal frequency of an illumination, utilising a custom-made swept source and a dual-axis galvanometric mirror in a common-path interferometer. Then, based on the principle of statistical optics, optical fields with respect to DSTCI were synthesised in both spatial and temporal frequency domains. Consequently, we quantified the speckle noise corresponding to various DSTCI and investigated the relationship between DSTCI and imaging quality.

## MATERIALS AND METHODS

For the systematic study on the effects of DSTCI on imaging quality, we measured the speckle noise in quantitative phase images with various DSTCI. We derived a decomposition of an optical field from an arbitrary low coherent illumination into optical fields with constituent spatiotemporal frequencies based on the principle of statistical optics. To synthesise optical fields using the decomposition with respect to various DSTCI, we measured optical fields with varying spatiotemporal frequency of the illumination.

In Figures 2(a-d), we depict the schematics of interferometric imaging systems with different DSTCI; (a) spatiotemporally coherent, (b) spatially coherent and temporally low coherent, (c) temporally coherent and spatially low coherent, and (d) spatiotemporally low coherent. In order to interpret the systems, we used two facts; one is the superposition principle in a linear system, and the other is the time averaging measurement in an optical wave regime due to slow detectors[33].

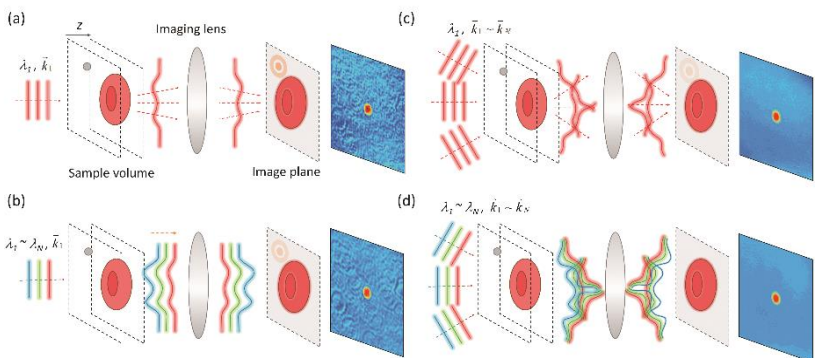

Figure 2. Schematics of interferometric imaging of a sample with a dust particle which is set apart from the sample along the optical axis. The right-side images are example phase images of a 3 μm–diameter polymethyl methacrylate bead with respect to the DSTCI. The degrees of coherence are (a) spatiotemporally coherent, (b) spatially coherent and temporally low coherent, (c) temporally coherent and spatially low coherent, and (d) spatiotemporally low coherent.

### Decomposition of an arbitrary low coherent illumination

In order to analyse an optical field from an arbitrary low coherent illumination, we decomposed the low coherent illumination into coherent illuminations with constituent spatiotemporal frequencies. The superposition principle enables the expression of the low coherent illumination as $E = \sum_{n}^{N} E_n = \mathrm{Re}\left[ \sum_{n}^{N} A_n e^{j\varphi_n} \times e^{j(\vec{k}_n \cdot \vec{r} - \omega_n t + \phi_n(\vec{k}_n, \omega_n))} \right]$, where $j$ is $\sqrt{-1}$, $A_n e^{j\varphi_n}$ is the complex amplitude of a wavefunction at the sample plane, $e^{j(\vec{k}_n \cdot \vec{r} - \omega_n t + \phi_n(\vec{k}_n, \omega_n))}$ is a propagating wave with monochromatic spatial $\vec{k}_n$, and temporal frequency $\omega_n$. $\phi_n(\vec{k}_n, \omega_n)$ denotes a relative phase among the propagating waves with different spatiotemporal frequencies, which is introduced to consider the low degree of coherence.

Since the relative phases among different frequencies of a low coherent light are randomly varying and unpredictable, a time averaging measurement leaves only terms independent of time and $\phi_n(\vec{k}_n, \omega_n)$.

To emulate interferometric imaging using a low coherent illumination, we consider the interference of the sample and reference beams. The two beams are assumed as a duplication of a low coherent light, and the sample beam contains diffracted light from a a sample. For the purpose of convenience, the optical field of the reference beam is 1, and is multiplied by a phase term $e^{j\theta}$ for phase retrieval;

$$S = \text{Re}\left[\sum_n^N A_n e^{j\varphi_n} \times e^{j(\vec{k}_n \cdot \vec{r} - \omega_n t + \phi_n(\vec{k}_n, \omega_n))}\right] \text{ and}$$

$$R = \text{Re}\left[\sum_n^N 1 \times e^{j(\vec{k}_n \cdot \vec{r} - \omega_n t + \phi_n(\vec{k}_n, \omega_n))} \times e^{j\theta}\right].$$

We first consider the low coherent illumination as a polychromatic plane wave which is the spatially coherent and temporally low coherent illumination, as shown in Figure 2(b). To start with the simplest case, we assume the illumination has two temporal frequencies $\omega_1$ and $\omega_2$. When the sample beam $S$ interferes with the reference beam $R$, the time averaging measurement can be written as

$$I = \langle |S+R|^2 \rangle_T = \sum_{n=1}^{2} \text{Re}\left[A_n^2 + 1 + A_n e^{j\varphi_n} \times e^{-j\theta} + A_n e^{-j\varphi_n} \times e^{j\theta}\right], \quad (1)$$

where $\langle \rangle_T$ means a time averaging measurement. The detailed derivation of Eq. (1) is presented in the **Supplementary**. Due to the time averaging measurement, the time dependent and the randomly varying relative phase terms vanish. By modulating the phase term $e^{j\theta}$, a phase retrieval method produces the resultant interference signal as $\sum_{n=1}^{2} A_n e^{j\varphi_n}$, which is a sum of optical fields from coherent illuminations with constituent temporal frequencies $\omega_1$ and $\omega_2$. Since the optical fields are summed independent of the propagating waves and the temporal frequency index $n$, Eq. (1) can be generalised to an arbitrary polychromatic illumination. Therefore, a resultant optical field from a polychromatic illumination is a sum of optical fields from coherent illuminations with constituent temporal frequencies $Ae^{j\varphi} = \sum_{n=1}^{N} A_n e^{j\varphi_n}$.

Next, we consider a monochromatic illumination passing through a rotating diffuser, which is the spatially low coherent and temporally coherent illumination [Figure 2(c)]; the rotating diffuser makes the relative phase relation among spatial frequencies randomly varying and unpredictable. Similar to the use of temporally low coherent illumination, we show that an optical field from a spatially low coherent illumination can be decomposed into a sum of optical fields from coherent illuminations with constituent spatial frequencies. To derive this result from the simplest case, the spatially low coherent illumination is assumed as having two spatial frequencies $\vec{k}_1$ and $\vec{k}_2$. Due to the use of a monochromatic illumination, time dependent terms vanish. Interference terms between different spatial frequencies remain; the interference terms between spatial frequencies, $e^{\pm j[(\vec{k}_1 - \vec{k}_2) \cdot \vec{r} + (\phi_1(\vec{k}_1) - \phi_2(\vec{k}_2))]}$, generates an interference pattern at the image plane. Since the rotating diffuser makes the relative phase difference $\phi_1(\vec{k}_1) - \phi_2(\vec{k}_2)$ randomly varying and unpredictable, the interference pattern becomes averaged out after time averaging measurement. As a result, the resultant interference signal becomes $\sum_{n=1}^{2} A_n e^{j\varphi_n}$ which can be generalised to various spatial frequencies similar to the result of temporally low coherent illumination $Ae^{j\varphi} = \sum_{n=1}^{N} A_n e^{j\varphi_n}$. In addition, due to the sum of optical fields with various spatial frequencies, the spatially low coherent illumination gives sample information beyond the numerical aperture limit, as shown in Supplementary Figure S1.

From the previous discussions for the use of temporally and spatially low coherent illuminations, an optical field from an arbitrary spatiotemporally low coherent illumination can be expressed as a sum of optical fields from coherent illuminations with constituent spatiotemporal frequencies [Figure 2(d)]. Therefore, using the decomposition, an optical field from an arbitrary low coherent illumination can be synthesised from measured optical fields with various spatiotemporal frequencies of illumination.

**Experimental measurements of optical fields with varying spatiotemporal frequency of illumination**

To synthesise optical fields using the decomposition with respect to various DSTCI, we measured optical fields with variation of the spatiotemporal frequency of the illumination by utilising a custom-made swept source and a dual-axis galvanometric mirror [Figure 3(a)].

Figure 3

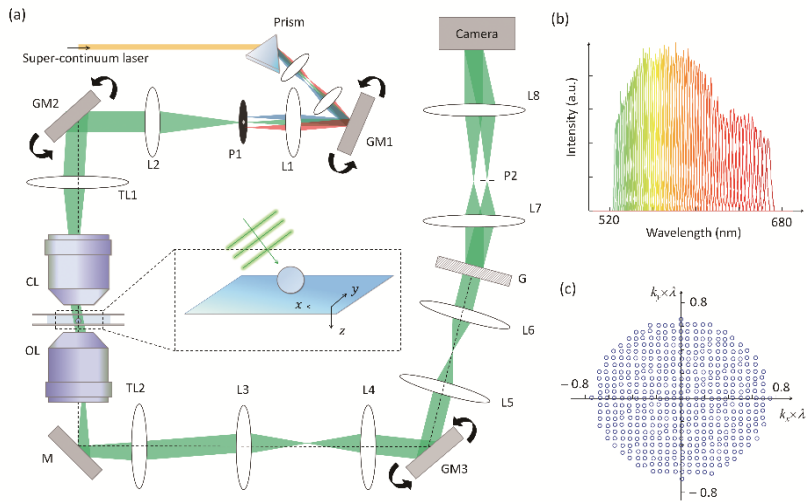

Figure 3. (a) Experimental setup. Common-path interferometry with varying spatiotemporal frequency of the illumination. The temporal frequency of the illumination is controlled by using the prism, the galvanometric mirror (GM1), and the pinhole (P1); the prism disperses an incident super-continuum laser, and the temporal frequency of light passing through P1 is controlled by the GM1. The dual-axis galvanometric mirror (GM2), which is located at the conjugate plane of a sample, controls the spatial frequency of the illumination. The other galvanometric mirror (GM3) compensates the tilted angle by the GM2 for common-path interferometric geometry. P1: pinhole (diameter = 100 μm); TL1: tube lens (f = 200 mm); CL: condenser lens, 0.9 NA; OL: objective lens, 1.42 NA; TL2: tube lens (f = 180 mm) L1: lens (f = 100 mm); L2: lens (f = 200 mm); L3,7: lenses (f = 75 mm); L4,8: lenses (f = 150 mm); L5,6: lenses (f = 100 mm); G: grating; P2: pinhole (diameter = 25 μm). (b) Temporal power spectra: centre wavelength is changed from 525 nm to 667 nm with 61 steps, the mean bandwidth is 4.6 nm. (c) Spatial frequencies controlled by the GM2 in (a) are uniformly scanned within the NA 0.75 with 441 steps.

For this work, we utilised a modified version of the experimental setup for hyperspectral optical diffraction tomography [Figure 3(a)][34]. In order to control the temporal frequency of the illumination, we utilised a custom-made wavelength-sweeping source. A prism (N-SF11 prism, PS853, Thorlabs Inc., NJ, USA) spectrally dispersed a polychromatic plane wave from a supercontinuum laser (SuperK compact, NKT photonics Inc., Birkerød, Denmark). The temporal frequency of light passing through the pinhole (P1) located at the Fourier plane of the prism was controlled by a galvanometric mirror (GM1, GVS011/M, Thorlabs Inc.). The centre wavelength of the illumination was scanned from 525 nm to 667 nm with 61 steps, and by changing the rotating angle of the GM1, the mean bandwidth was 4.6 nm. The temporal power spectra measured by a spectrometer (HR4000, Ocean Optics Inc.) is presented in Figure 3(b).

The spatial frequency of the illumination was controlled by tilting the angle of a dual-axis galvanometric mirror (GM2, GVS012/M, Thorlabs Inc.) located at the conjugate plane of the image plane. The spatial frequency was uniformly scanned over lateral directions within the NA = 0.75 with 441 steps [Figure 3(c)]. By combining the scanning of the spatial and temporal frequency of the illumination, we independently controlled the spatiotemporal frequency with $61 \times 441 = 26901$ conditions. The plane wave with the controlled spatiotemporal frequency was further demagnified by using a tube lens (TL1, $f$ = 200 mm) and a condenser lens (CL, UPLFLN 60×, NA = 0.9, Olympus Inc., Tokyo, Japan), and impinged onto a sample, as depicted in the inset of Figure 3(a).

For interferometric imaging with varying spatiotemporal frequency of the illumination, we utilised a common-path interferometry[35-39]. The light diffracted from a sample was collected by an objective lens (OL, UPLSAPO 60×, oil immersion, NA = 1.42, Olympus Inc.). The other galvanometric mirror (GM3, GVS012/M, Thorlabs Inc.) compensated the scanning angle tilted by the GM2 so that the optical axis after the GM3 remained fixed over various spatial frequencies. A phase grating (G, 92 grooves/mm, #46-072, Edmund Optics Inc., NJ, USA) replicated the optical field of a sample into many diffraction orders. The pinhole, located at the Fourier plane of the grating, passed the 1st diffraction beam and spatially filtered the 0th diffraction beam so that it was converted into a plane wave after the lens L8, which was used as the reference arm. In order to avoid dispersion of the grating, we selected the 0th diffraction beam for spatial filtering. Then, the sample field and the filtered plane waves interfered at the camera plane and generated spatially modulated holograms. The holograms were recorded by a scientific complementary metal-oxide semiconductor camera (Neo sCMOS, ANDOR Inc., Northern Ireland, UK). Optical fields were retrieved from the measured holograms by implementing the field retrieval algorithm based on Fourier transform[40, 41].

### Sample preparation
We used polymethyl methacrylate (PMMA) microspheres with a diameter of 3 μm ($n$ = 1.4934 at $\lambda$ = 532 nm, Sigma-Aldrich Inc., MO, USA) immersed in index-matching oil ($n$ = 1.4227 at $\lambda$ = 532 nm, Cargille Laboratories, NJ, USA). In order to separate aggregated microspheres, the microspheres in the immersion oil were sonicated and sandwiched between coverslips before measurements.

## RESULTS AND DISCUSSION
For the quantitative study on the effect of DSTCI on imaging quality, we quantify DSTCI and the speckle noise by the spatial and temporal coherence lengths and the standard deviation of a phase image in the absence of a sample, respectively. We synthesise optical fields using the decomposition with respect to various DSTCI. Then, the speckle noise is measured from the synthesised phase images and plotted with respect to the spatial and temporal coherence lengths calculated from power spectra of illuminations.

## Synthesis of optical fields with respect to various DSTCI of the illumination

In order to explore the effect of DSTCI on imaging quality, optical fields with respect to various DSTCI are synthesised by a sum of measured optical fields with constituent spatiotemporal frequencies [MATERIALS AND METHODS]. The degrees of spatial and temporal coherence are individually controlled with 13 and 31 steps, respectively, and thereby the whole DSTCI is controlled with 13×31 = 403 steps. The representative synthesised optical fields of a PMMA microsphere are shown in Figure 4. The schematics and power spectra visualise the controlled DSTCI. The insets depict magnified images.

### Figure 4

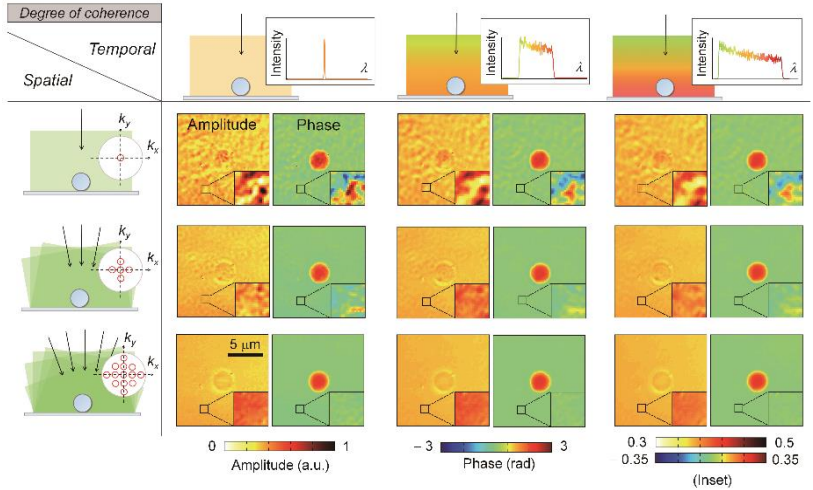

Figure 4. Representative optical field images of a PMMA microsphere illuminated with various DSTCI. The inset shows magnified images. The degree of spatial and temporal coherence of the illumination decreases along the row and column, respectively.

## Spatial and temporal coherence lengths of the synthesised illuminations

To quantitatively investigate, we quantify DSTCI by calculating the spatial and temporal coherence lengths from spatial and temporal autocorrelation functions (ACF), respectively. We obtain the spatial and temporal ACF from the Fourier transformation (FT) of illumination power spectra using the Wiener–Khinchin theorem, $\text{ACF}[f(x)] = \text{FT}\left(|F(k)|^2\right)$ [Figures 5(a-b) and (d-e)]. For visualisation purposes, the spatial power spectra are presented azimuthally averaged, and $s$ denotes the distance from the origin.

From the obtained spatial ACFs, the spatial coherence length ($l_s$), which is defined as peak to first zero distance of a spatial ACF, can be expressed by the spatial frequency bandwidth ($\Delta k_s$) as $l_s = 1.22/\Delta k_s$ [31, 33]. Similarly, the temporal coherence length ($l_t$), which is defined as full–width half maximum of a temporal ACF, can be expressed by the centre wavelength ($\lambda_0$), a refractive index of the medium ($n_{medium}$), and wavelength bandwidth ($\Delta\lambda$) as $l_t = \lambda_0^2/n_{medium}\Delta\lambda$ [1, 33]. The obtained spatial and temporal coherence lengths with respect to DSTCI are presented in Figures 5(c) and (f) with the theoretically predicted curve.

### Figure 5

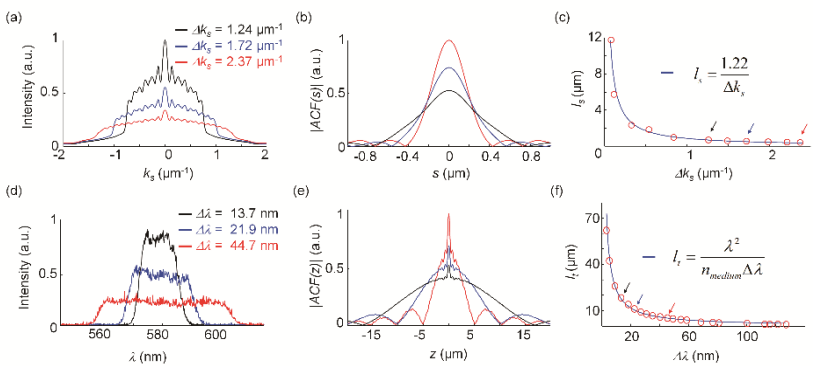

Figure 5. Spatial and temporal power spectra, autocorrelation functions (ACF), and coherence lengths with respect to bandwidths. (a) Representative azimuthally averaged spatial power spectra of the illuminations. Solid black, blue, and red lines correspond to the spatial frequency bandwidth ($\Delta k_s$) = 1.24, 1.72, and 2.37 μm$^{-1}$, respectively. (b) The absolute of the spatial ACF corresponding to (a). (c) Spatial coherence lengths ($l_s$) as a function of spatial frequency bandwidths. The fitted curve is $l_s = 1.22/\Delta k_s$. The spatial power spectra in (a) are marked as black, blue, and red arrows, respectively. (d) Representative temporal power spectra with the centre wavelength of 579 nm. Solid black, blue, and red lines correspond to the wavelength bandwidth ($\Delta\lambda$) of 13.7, 21.9, and 44.7 nm, respectively. (e) The absolute of the temporal ACF corresponding to (d). (f) Temporal coherence lengths ($l_t$) with respect to wavelength bandwidths. The fitted curve is $l_t = \lambda^2/(n_{medium}\Delta\lambda)$, where $n_{medium}$ is the refractive index of the medium. The temporal power spectra in (d) are marked as black, blue, and red arrows, respectively.

## The effects of DSTCI on imaging quality

In order to systematically analyse the effects of DSTCI on imaging quality, we measure the speckle noise with respect to spatial and temporal coherence lengths of the illumination. For quantitative study, we quantify the speckle noise as the standard deviation of a phase image in the absence of a sample[12]. For statistically relevant analysis, a phase image in the absence of a sample is partitioned into 96 regions, and the speckle noise with respect to spatial and temporal coherence length is specified by averaging the standard deviations of each partitioned region.

Figure 6(a) presents the relationship between the speckle noise ($\sigma_\varphi$) and the spatial and temporal coherence lengths ($l_s$, $l_t$) in a logarithmic scale. The black dots represent the measured speckle noise, and the surface plot is obtained by linear interpolation of the measured results for visualisation purposes. It clearly shows that the speckle noise decreases with decreasing the spatial and temporal coherence length. The speckle noise decreases from 120 mrad at the longest $l_s$ and $l_t$, to 8.3 mrad at the shortest $l_s$ and $l_t$. Furthermore, by comparing the vertices of the surface, a synergistic effect between spatially and temporally low coherent illuminations can be verified; the speckle noise is significantly reduced by both spatially and temporally low coherent illumination as opposed to either spatially or temporally low coherent illumination.

For further analysis, cross-sectional views of Figure 6(a) at the $l_t$ = 63.3 μm and $l_s$ = 11.8 μm are shown in Figure 6(b-c) in a logarithmic scale, respectively. As shown in Figure 6(b), the speckle noise increases from 17 mrad to 120 mrad as the spatial coherence length increases from 0.408 μm to 11.8 μm with the mean slope 0.57. In contrast, the speckle noise increases from 70 mrad to 120 mrad as the temporal coherence length increases from 1.4 μm to 63.3 μm with the gradually decreasing slope from 0.22 to 0.08 [Figure 6(c)]. The slopes provide quantitative measures of the speckle noise reduction by either spatially or temporally shortened coherence length. A comparison of the slopes indicates that decreasing the degrees of spatial coherence reduces the speckle noise 2.6 – 7.1 times greater than decreasing the degrees of temporal coherence.

In order to compare the result with a coherent illumination, we also measure the speckle noise with a spatiotemporally coherent illumination in the same setup. We utilise a diode-pumped solid-state laser (Samba™, Cobolt Inc., Sweden) with a coherence length longer than 10 m. The speckle noise with the spatiotemporally coherent illumination is 220 mrad, which is marked with the red circle in Figure 6(c).

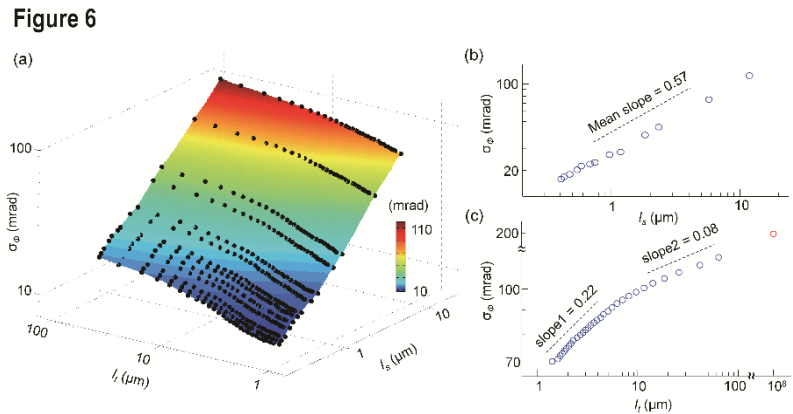

Figure 6. Speckle noise quantified as the spatial phase standard deviation ($\sigma_\varphi$) with respect to the spatial and temporal coherence lengths ($l_s$, $l_t$). (a) Measured $\sigma_\varphi$ are presented by black dots corresponding to the spatial and temporal coherence lengths, and the surface plot is obtained by the linear interpolation from the measured results for visualisation purposes. (b-c) A cross-sectional view of (a) at the $l_t$ = 63.3 μm and $l_s$ = 11.8 μm, respectively. The red circle marker in (c) corresponds to a coherent illumination.

The result provides the synergistic effect and the comparison of reducing the speckle noise between the spatially and temporally low coherent illuminations. The effect of DSTCI on the speckle noise can be interpreted with variations of correlated sample volume. The speckle noise, which originates in defocused samples or dust particles, is averaged out by the sum of propagated waves with the diversity of the spatiotemporal frequencies, so that the correlated sample volume among the propagating waves shrinks. In addition, the spatially low coherent illumination has less correlated volume than the temporally low coherent illumination, which makes the difference of speckle noise reduction by spatially and temporally low coherent illuminations.

## CONCLUSIONS

In sum, we systematically investigated the relationship between DSTCI and imaging quality with the same imaging condition. For this purpose, we showed that an optical field from an arbitrary low coherent illumination can be decomposed into optical fields from coherent illuminations with the constituent spatiotemporal frequencies. We then measured optical fields using a custom-made swept source common-path interferometer with varying spatiotemporal frequency of the illumination. By exploiting the decomposition, optical fields with respect to various DSTCI were synthesised from the measured optical fields. For quantitative study, DSTCI and the speckle noise were quantified by spatial and temporal coherence lengths and standard deviation of a phase image in the absence of a sample, respectively. The results reveal the synergistic effect and the comparative effects on the speckle noise reduction between the spatially and temporally low coherent illuminations. Although the present work discusses the effects of DSTCI on interferometric imaging, the method and analysis presented in this work can be readily applied to intensity imaging because the current discussions are based on optical field information.

This study can provide an important reference when selecting an appropriate degree of spatiotemporal coherence of an illumination. Furthermore, through this study, the intrinsic nature of interferometric imaging using an arbitrary low coherent illumination becomes better understood. We believe that the presented methods and results can be implemented in various interferometric imaging systems for characterising and improving imaging quality.

## Acknowledgements
This work was supported by KAIST, and the National Research Foundation of Korea (2015R1A3A2066550, 2014K1A3A1A09063027, 2014M3C1A3052567) and Innopolis foundation (A2015DD126).

### Competing Financial Interests
Prof. Park has financial interest in Tomocube Inc., a company that commercializes optical diffraction tomography and quantitative phase imaging instruments and is one of the sponsors of the work.